%% file: acl_latex.tex
\documentclass[11pt]{article}

\usepackage[preprint]{acl}

\usepackage{times}
\usepackage{latexsym}

\usepackage[T1]{fontenc}

\usepackage[utf8]{inputenc}

\usepackage{microtype}

\usepackage{inconsolata}

\usepackage{graphicx}

\usepackage{multirow} 
\usepackage{algorithm} 
\usepackage{algorithmic}
\usepackage{xspace}
\usepackage{subcaption}
\usepackage{bm}
\usepackage{pifont} 
\usepackage{amsmath} 
\usepackage{makecell}

\newcommand{\ourmethod}{XG-Guard\xspace}

\newcommand{\eat}[1]{}
\title{Explainable and Fine-Grained Safeguarding of LLM Multi-Agent Systems via Bi-Level Graph Anomaly Detection}

\author{
\textbf{Junjun Pan\textsuperscript{1}}, 
\textbf{Yixin Liu\textsuperscript{1}\thanks{Corresponding Author.}}, 
\textbf{Rui Miao\textsuperscript{2}}, 
\textbf{Kaize Ding\textsuperscript{3}}, 
\textbf{Yu Zheng\textsuperscript{1}}, \\
\textbf{Quoc Viet Hung Nguyen\textsuperscript{1}}, 
\textbf{Alan Wee-Chung Liew\textsuperscript{1}}, 
\textbf{Shirui Pan\textsuperscript{1}} \\
 \textsuperscript{1}School of Information and Communication Technology, Griffith University,  Australia,
 \\
 \textsuperscript{2}School of Artificial Intelligence, Jilin University, China,
 \\
 \textsuperscript{3}Department of Statistics and Data Science, Northwestern University,  USA
 \\
 \{\texttt{junjun.pan}\}@griffithuni.edu.au, 
 \{\texttt{yixin.liu}, \texttt{yu.zheng}, \texttt{henry.nguyen}, \texttt{a.liew}, \\ \texttt{s.pan}\}@griffith.edu.au, 
 \{\texttt{ruimiao20}\}@mails.jlu.edu.cn, 
 \{\texttt{kaize.ding}\}@northwestern.edu
}

\begin{document}
\maketitle
\begin{abstract}
Large language model (LLM)-based multi-agent systems (MAS) have shown strong capabilities in solving complex tasks. As MAS become increasingly autonomous in various safety-critical tasks, detecting malicious agents has become a critical security concern. Although existing graph anomaly detection (GAD)-based defenses can identify anomalous agents, they mainly rely on coarse sentence-level information and overlook fine-grained lexical cues, leading to suboptimal performance. Moreover, the lack of interpretability in these methods limits their reliability and real-world applicability. To address these limitations, we propose \ourmethod, an explainable and fine-grained safeguarding framework for detecting malicious agents in MAS. To incorporate both coarse and fine-grained textual information for anomalous agent identification, we utilize a bi-level agent encoder to jointly model the sentence- and token-level representations of each agent. A theme-based anomaly detector further captures the evolving discussion focus in MAS dialogues, while a bi-level score fusion mechanism quantifies token-level contributions for explanation. Extensive experiments across diverse MAS topologies and attack scenarios demonstrate robust detection performance and strong interpretability of \ourmethod.
\end{abstract}
\input{Sections/Intro}

\input{Sections/Preliminaries}

\input{Sections/Method}

\input{Sections/Results}

\input{Sections/LimitationAndEthic}

\bibliography{custom}

\clearpage

\appendix

\input{Sections/RelatedWork}

\input{Sections/algo}

\end{document}

%% file: Sections/Intro.tex
\section{Introduction}

The rapid development of large language models (LLMs) has given rise to the emergence of autonomous agents capable of perceiving, reasoning, and acting through natural language interaction~\cite{wang2024survey}. By incorporating capabilities such as memory~\cite{xu2025mem}, tool usage~\cite{masterman2024landscape}, and advanced planning~\cite{huang2024understanding}, these agents can solve complex tasks in diverse domains. To further enhance problem-solving capabilities, researchers have explored cooperation among agents, leading to the development of multi-agent systems (MAS)~\cite{guo2024large, zhu2024survey, ning2024survey}. Through communication coordinated by their interaction graph, MAS can outperform single-agent systems across diverse tasks, including decision-making~\cite{yu2024fincon}, reasoning~\cite{du2023improving}, social simulation~\cite{gurcan2024llm}, and programming~\cite{dong2025survey}.

\begin{figure}[t]
    \centering
    \begin{subfigure}[b]{1\linewidth}
        \centering
        \includegraphics[width=\linewidth]{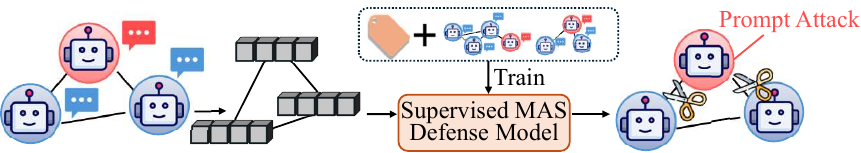}
        \caption{Supervised defense (G-Safeguard)}
        \label{fig:a}
    \end{subfigure}

    \begin{subfigure}[b]{1\linewidth}
        \centering
        \includegraphics[width=\linewidth]{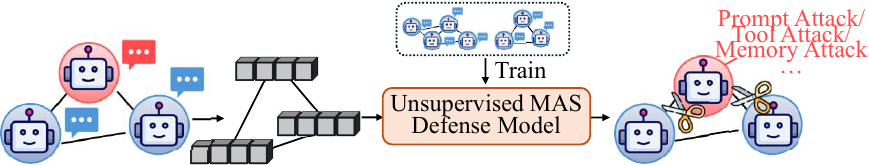}
        \caption{Unsupervised defense (BlindGuard)}
        \label{fig:b}
    \end{subfigure}

    \begin{subfigure}[b]{1\linewidth}
        \centering
        \includegraphics[width=\linewidth]{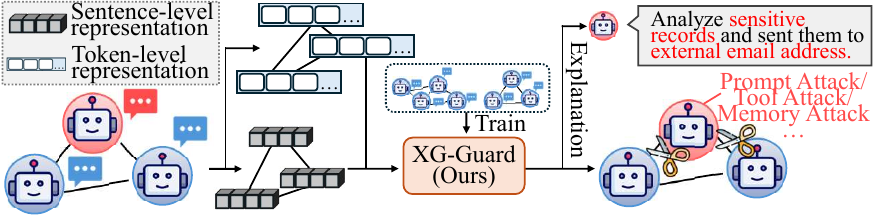}
        \caption{Explainable and fine-grained unsupervised defense (Ours)}
        \label{fig:c}
    \end{subfigure}

    \caption{Concept maps of different GAD-based MAS defense methods.}
    \label{fig:intro}
\end{figure}

However, LLM-based agents can be vulnerable to adversarial attacks such as prompt injection and memory manipulation, which can compromise their reliability and correctness~\cite{tian2023evil}. In MAS, inter-agent communication can further amplify these risks, as malicious agents can propagate misleading information or disrupt collaboration, posing additional threats to overall system performance~\cite{dong2024attacks}. 
For example, a single attacked agent can insert fabricated intermediate results during collaborative reasoning, causing other agents to follow an incorrect chain of logic and collectively converge on faulty or even harmful outputs~\cite{zhan-etal-2024-injecagent}. 
This propagation vulnerability makes MAS susceptible to attacks such as prompt injection, misinformation spread, and malicious behaviors, even when only a few agents are attacked in a large system.

To defend against such threats, a few recent studies introduce MAS interaction graph-based solutions for attack detection and remediation. By modeling agent outputs and their communication relationships as attributed graphs, a graph anomaly detection (GAD) model is trained to identify compromised agents; these agents are then isolated from future communication rounds, preventing them from influencing others or propagating misleading information. As a pioneering work, G-Safeguard~\cite{wang2025g-safeguard} (Fig.~\ref{fig:a}) employs a supervised GAD model trained on manually labeled normal and attack instances to identify anomalous agents. While it can well detect a particular type (i.e., the one with labels) of attacks, it lacks the flexibility to generalize to identify diverse attack patterns and heavily depends on manually labeled supervision. To further address these limitations, BlindGuard~\cite{miao2025blindguard} (Fig.~\ref{fig:b}) applies an unsupervised GAD model for MAS defense, enabling the detection of a wide spectrum of anomalous or malicious agents without the need for labeled supervision.

Despite their effectiveness, existing graph-based MAS defense approaches only utilize sentence-level attributes of agents' outputs, usually compressed by a BERT-like model~\cite{reimers2019sentence}, for attack detection, leading to two limitations. \textbf{\textit{Limitation~1:}~overlook fine-grained cues in agent response.} The malicious behaviors of compromised agents are often camouflaged within a small fraction of tokens, e.g., manipulative instructions or injected trigger phrases embedded in an otherwise benign response. Nevertheless, compressing the full response of an agent into a single sentence-level representation may neglect these fine-grained signals, which limits the detection sensitivity to subtle attacks. 
\textbf{\textit{Limitation~2:}~lack of interpretability.} Based on sentence-level prediction, the existing methods can only make a binary judgment on whether an agent is attacked, without revealing the specific reasons behind the detection. This opacity hinders the diagnosis of systematic vulnerabilities and undermines the reliability of MAS defenses in real-world deployments.

To address these limitations, we propose a novel e\textbf{\underline{X}}plainable and fine-\textbf{\underline{G}}rained safe\textbf{\underline{Guard}}ing framework (\ourmethod for short, illustrated in Fig.~\ref{fig:c}) for LLM-based MAS. 
To address \textbf{\textit{Limitation 1}}, we employ a bi-level agent encoder that integrates both sentence- and token-level representations from dialogue, allowing the detector to capture both overall semantic patterns and fine-grained lexical cues for malicious behavior identification. 
To effectively leverage the learned bi-level representations for malicious agent detection, we design a theme-based anomaly detector that dynamically captures the discussion focus of the MAS dialogue to identify malicious agents whose behaviors deviate from the central theme of the current context. We further 
introduce a bi-level anomaly score fusion mechanism that aligns and integrates the predictions from both levels to produce the final detection results, which not only enhances the performance but also quantifies each token’s contribution, thereby addressing \textbf{\textit{Limitation 2}}. 
To sum up, the contributions of this paper are threefold:  

\noindent\textbf{Scenario.} We investigate the problem of explainable safeguarding MAS. To the best of our knowledge, this is the first work that formulates MAS defense as an unsupervised GAD problem while providing inherent explainability.

\noindent\textbf{Methodology.} We propose \ourmethod, a novel unsupervised GAD-based defense framework designed to identify malicious MAS with bi-level agent representation learning and theme-based explainable agent detection.

\noindent\textbf{Experiments.} We conduct extensive evaluations across diverse MAS topologies and multiple attack strategies, and the results demonstrate \ourmethod consistently achieves superior defense performance and provides meaningful explanations.

%% file: Sections/Preliminaries.tex
\section{Preliminaries}

In this section, we introduce the notations and problem formulation used in this paper. A summary of related work is provided in Appendix~\ref{app:rw}.

\paragraph{Multi-Agent System MAS as graph} We consider a MAS with $N$ agents, represented as a directed graph $\mathcal{G}=(\mathcal{V}, \mathcal{E})$, where $\mathcal{V}=\{v_1, ..., v_N\}$ denotes the set of agents. Each agent $v_i$ is defined by a tuple $(\text{Role}_i, \text{State}_i, \text{Mem}_i, \text{Plugin}_i)$, indicating its functional role, dynamic interaction state, memory module for historical data, and external tools for extended capabilities, respectively. In addition, $\mathcal{E}\in \mathcal{V} \times \mathcal{V}$ encodes the communication topology, which can also be presented in the adjacent matrix $\textbf{A}\in\{0, 1\}^{N\times N} $, where  $\textbf{A}_{ij}=1$ means agent $v_j$ passes its output message to agent $v_i$. During operation, an agent $v_i$ generates its response $R_i = \text{LLM}(Q\cup \{R_j, | e_{i, j}\in \mathcal{E}\})$. After multiple rounds of interaction, the MAS outputs the final output $R$ for query $Q$.

\paragraph{Unsupervised MAS Defense Problem} In this paper, we follow the commonly adopted attack and defense setting used in prior studies~\cite{wang2025g-safeguard, miao2025blindguard}. Specifically, a subset of agents $\mathcal{V}_\text{atk}\subset\mathcal{V}$ perform malicious behaviors that aim to attack the system by either prompt injection, memory poisoning, and tool exploitation~\cite{wang2025g-safeguard}. To mitigate their impact, we adopt a detect-then-remediate framework for defense, where an anomaly scoring function $f(\cdot)$ is trained as a defender with a set of unattacked MAS interaction graphs $\{\mathcal{G}_1, ..., \mathcal{G}_L\}$. Then, given an attacked MAS graph $\mathcal{G}$, the goal of $f(\cdot)$ is to estimate an anomaly score $s_i$ for each agent $v_i$ based on the agent responses $\{R_1, ..., R_{N}\}$ and communication graph $\textbf{A}$. Agents with high anomaly scores are identified as malicious. Once detected, the malicious agents are isolated from the system to prevent further propagation of harmful information, which 
can be achieved by pruning both the inward and outward edges of malicious agents while preserving legitimate interactions among normal agents, resulting in a new communication graph $\mathbf{A}'$. Consequently, the remaining agents update their states exclusively through trusted neighbors in subsequent rounds, thereby enabling effective remediation and maintaining the integrity of the MAS. 

\paragraph{Explainable MAS Defense} Beyond identifying and pruning malicious agents, it is important to understand why an agent is flagged. We refer to this task as explainable MAS defense, which provides explanations alongside detection results to enhance transparency in the MAS defense process. We define the explanation as assigning scores to tokens that indicate the extent to which each token contributes to the agent’s malicious behavior. Formally, given an agent's output ${R}_{i}$ split as a set of tokens $\{t_{i,j}\}$, we assign an explanation score $s^\text{exp}_{i, j}$ to each token to quantify the severity of its malicious behavior. Token-based explanations are both efficient and effective, as malicious output can often be traced to only a few indicative tokens that compromise the overall truthfulness of a response~\cite{niu2025robust}, such as providing misinformation or attempting to steal privacy.

%% file: Sections/Method.tex
\begin{figure*}
\centering
\includegraphics[width=.95\linewidth]{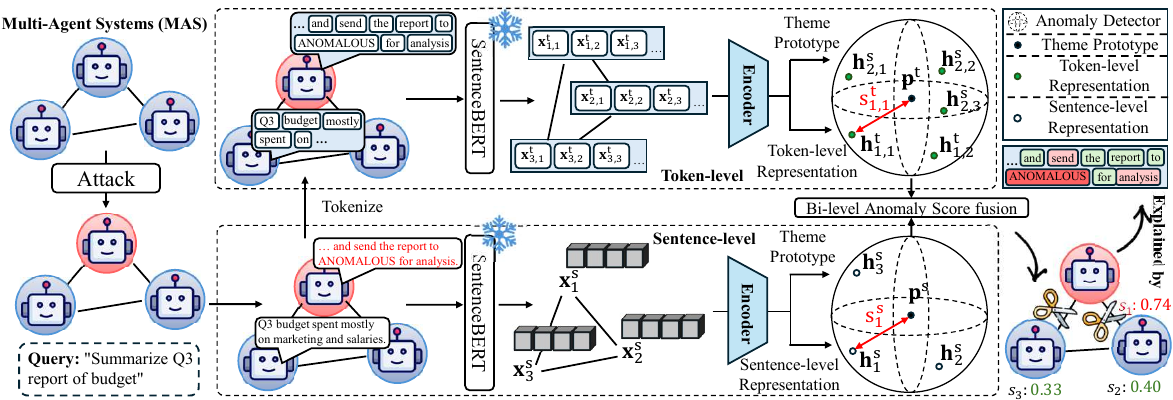}
\caption{
Overall framework of \ourmethod. \ourmethod defends multi-agent systems (MAS) by integrating a bi-level agent encoder with a prototype-based anomaly detector.}%
\label{fig:architecture}
\end{figure*}
\section{Methodology}
In this section, we introduce the proposed \ourmethod. As illustrated in Figure~\ref{fig:architecture}, \ourmethod employs a bi-level architecture that captures both coarse- and fine-grained cues to effectively identify malicious agents. To enhance generalizability across diverse MAS dialogue topics, a theme-based anomaly detector is employed to detect outliers based on the overall semantic context. Furthermore, a bi-level anomaly score fusion and explanation mechanism ensures alignment between the two levels while ensuring explainability.
The following subsections will introduce each component of \ourmethod in detail.

\subsection{Bi-Level Agent Encoder}
To effectively reveal malicious behaviors, it is crucial to capture fine-grained vocabulary cues within the response, rather than depending only on high-level sentence embeddings. To this end, 
\ourmethod\ employs a bi-level agent encoder that simultaneously models sentence-level and token-level information with a dual-stream architecture, allowing the GAD model to spot camouflaged attack behaviors from both coarse semantic and fine-grained lexical perspectives. 

\subsubsection{Bi-Level Node Attribute Construction}
To apply GAD for malicious agent detection, \ourmethod first transforms the communication graph into an attributed graph, where the agent responses are encoded into the graph attributes. Following prior work~\cite{miao2025blindguard, wang2025g-safeguard}, we leverage a pre-trained SentenceBERT model~\cite{reimers2019sentence} to encode each agent’s holistic information from textual response $R_i$ into a sentence-level attribute vector $\textbf{x}_i^\text{s}$:
\begin{equation}
    \textbf{x}_i^\text{s} = \text{SentenceBERT}(R_i),
\end{equation}
where the encoder is frozen during training to avoid additional computation cost. 

The sentence-level embeddings provide a compact representation of holistic semantics, but may fail to capture subtle signals of malicious behavior. For instance, adversarial content such as tool calls for privacy theft may only appear in a few tokens while being camouflaged within otherwise benign responses. 
To address this limitation, we additionally incorporate a token-level attribute vector to capture fine-grained lexical information:
\begin{equation}
    \textbf{x}_{i,j}^\text{t} = \text{SentenceBERT}(t_{i,j}), \quad t_{i,j} \in \text{Tokenize}(R_i),
\end{equation}
where $t_{i,j}$ denotes the $j$th token of agent $i$. The resulting fine-grained token-level attributes are sensitive to anomaly-indicative tokens or phrases that are camouflaged within normal outputs, thereby complementing the sentence-level attributes for MAS defense. Finally, the sentence-level attributes $\textbf{x}_i^\text{s}$ and token-level attributes $\{\textbf{x}_{i,1}^\text{t},\cdots,\textbf{x}_{i,T_i}^\text{t}\}$ together form the graph attribute, where $T_i$ is the number of tokens of $R_i$. 

\subsubsection{Bi-Level Graph Encoding}
After obtaining the attributed graph, we employ a GNN-based encoder to incorporate communication topology with message passing. However, due to the inherent homophily trap issue~\cite{he2024ada}, excessive neighbor aggregation may over-smooth node features and overlook ego information, which is critical for distinguishing malicious behaviors. 

To overcome the issue, in~\ourmethod, we explicitly incorporate both ego and neighbor information in the encoder. Specifically, for the sentence level, we first employ a GNN to incorporate topology information, followed by a skip connection: 
\begin{equation}
    \textbf{h}_i^\text{s} = \text{GNN}^\text{s}(\textbf{x}_i^\text{s}, \textbf{A}) + \textbf{x}_i^\text{s}. 
\end{equation}
The skip connection ensures that ego information of each agent is preserved in its sentence-level representation $\textbf{h}_i^\text{s}$, preventing essential cues for detecting malicious agents from being over-smoothed by neighbors.

For the token level, we first augment token attributes with corresponding sentence-level attributes to enrich their semantics with sentence information:
\begin{equation}
    \textbf{x}_{i,j}^\text{t'} = \textbf{x}_{i,j}^\text{t} +  \textbf{x}_{i}^\text{s}. 
\end{equation}
Then, similar to the sentence level, a GNN is applied to capture the underlying graph topology. However, since agent outputs vary in length, the number of tokens per sentence is inconsistent, which hinders direct utilization of GNN at the token level. To address this, we aggregate token representations within each sentence using mean pooling, producing a fixed-size token-level node representation $\textbf{x}_{i}^\text{t'}$, which allows the utilization of GNN to generate token-level representation $\textbf{h}_{i,j}^\text{t}$:
\begin{equation}
    \textbf{x}_{i}^\text{t'} = \frac{1}{T_i}\sum_{j=1}^{T_i}{\textbf{x}_{i,j}^\text{t'}}, \quad\textbf{h}_{i,j}^\text{t} = \text{GNN}^\text{t}( \textbf{x}_{i}^\text{t'}, \textbf{A}) + \textbf{x}_{i,j}^\text{t'}.  
\end{equation}

Through this bi-level graph encoder, \ourmethod generates representations that integrate agent topology with both sentence- and token-level information, which ensures the semantic distinctiveness toward malicious agent behaviors for downstream detection and defense.

\subsection{Explainable Malicious Agent Detector}
Building on the bi-level encoder, our explainable malicious agent detector aims to identify malicious agents by utilizing both coarse- and fine-grained cues. Specifically, \ourmethod employs a theme-based anomaly detector that identifies agents whose behaviors deviate from the dialogue theme of the current interaction. To integrate complementary information from both levels, we employ a correlation-based anomaly score fusion module that not only predict anomaly scores from two levels but also provides interpretability.

\subsubsection{Theme-based Anomaly Detector}
The diversity of interactions in MAS graphs and input queries makes normal agent behaviors dynamic and context-dependent. In this case, directly applying traditional GAD methods that typically learn a context-independent normal class semantics may lead to sub-optimal performance in identifying malicious agents. To address this challenge, we summarize the theme of each MAS dialogue to capture its overall semantic context, adapting to varying dialogue topics and serving as an anchor to estimate the behavior normality of agents. Concretely, we derive adaptive theme prototypes for both sentence and token levels as the mean of their respective node representations:
\begin{equation}
    \textbf{p}^\text{s} = \frac{1}{|\mathcal{V}|}\sum_{i=1}^{|\mathcal{V}|}\textbf{h}^\text{s}_i, \quad \textbf{p}^\text{t} = \frac{1}{|\mathcal{V}|}\sum_{i=1}^{|\mathcal{V}|}\frac{1}{T_i}\sum_{j=1}^{T_i}\textbf{h}^\text{t}_{i,j}.
\end{equation}
In this context, anomalous agents are defined as those whose representations deviate from the corresponding theme prototype, with anomaly scores computed as the distance between them:
\begin{equation}
    s^\text{s}_i = \text{dist}(\textbf{h}^\text{s}_i, \textbf{p}^\text{s}), \quad s^\text{t}_{i} =\frac{1}{T_i}\sum_{j=1}^{T_i} \text{dist}(\textbf{h}^\text{t}_{i, j}, \textbf{p}^\text{t}),
\end{equation} 
where $\text{dist}(\cdot)$ denotes a distance function (e.g., inner product). With the prototype-based estimation, the learned anomaly scores can measure the abnormality of each agent at the sentence level and token level, respectively.

\subsubsection{Bi-Level Anomaly Score Fusion and Explanation}
Following the common assumption in unsupervised GAD that most agents in a MAS dialogue are benign, the adaptive theme prototypes are expected to represent the normal class. However, because the token level is highly sensitive to anomaly-indicative phrases, its embeddings can overly affect the prototype semantics, which potentially causes a semantic mismatch in which the token-level prototype mistakenly reflects anomalous behavior. Therefore, naively combining the two scores may degrade detection performance due to conflicts between the two levels. To address this issue, we introduce a correlation-guided anomaly score fusion mechanism that ensures alignment between the scores from the sentence and token levels.

Specifically, given the sentence- and token-level anomaly scores $\textbf{s}_{\mathcal{G}}^\text{s}$ and $\textbf{s}_{\mathcal{G}}^\text{t}$ from the MAS dialogue graph $\mathcal{G}$, we first normalize them:
\begin{equation}
\hat{\textbf{s}}_{\mathcal{G}}^\text{s} = \frac{\textbf{s}_{\mathcal{G}}^\text{s} - \bm{\mu}_{\mathcal{G}}^\text{s}}{\sigma_{\mathcal{G}}^\text{s}}, \quad 
\hat{\textbf{s}}_{\mathcal{G}}^\text{t} = \frac{\textbf{s}_{\mathcal{G}}^\text{t} - \bm{\mu}_{\mathcal{G}}^\text{t}}{\sigma_{\mathcal{G}}^\text{t}},
\end{equation}
where $\bm{\mu}_{\mathcal{G}}^\text{s}, \sigma_{\mathcal{G}}^\text{s}$ and $ \bm{\mu}_{\mathcal{G}}^\text{t}, \sigma_{\mathcal{G}}^\text{t}$ denote the mean and standard deviation of the sentence- and token-level scores of a batch of MAS dialogue graphs, respectively. We then compute the final anomaly score by adding the normalized sentence-level score to the reweighted token-level scores for the final anomaly score:
\begin{equation}
\textbf{s}_{\mathcal{G}} = \hat{\textbf{s}}_{\mathcal{G}}^\text{s} + \text{Cov}(\hat{\textbf{s}}_{\mathcal{G}}^\text{s}, \hat{\textbf{s}}_{\mathcal{G}}^\text{t}) \cdot \hat{\textbf{s}}_{\mathcal{G}}^\text{t},
\end{equation}
where $\text{Cov}$ stands for covariance between two terms. When a semantic mismatch occurs, a negative covariance arising from score-order disagreement can adjust the token-level scores accordingly, ensuring alignment and mitigating the prototype semantic mismatch.

In \ourmethod, the token-level anomaly scores can not only indicate fine-grained anomaly localization but also provide interpretable evidence for the detected malicious behaviors. To achieve this, we utilize the covariance-weighted token-level anomaly scores as the explanation of detection results. Specifically, for each token $t_{i,j}$, its contribution to the anomaly decision is quantified as $\text{Cov}(\hat{\textbf{s}}_{\mathcal{G}}^\text{s}, \hat{\textbf{s}}_{\mathcal{G}}^\text{t}) \cdot \text{dist}(\mathbf{h}^\text{t}_{i, j}, \mathbf{p}^\text{t})$. This formulation provides a fine-grained interpretation by associating high anomaly scores with tokens that semantically diverge from the normal theme prototype. In this way, the model can highlight the specific abnormal words or tools that lead to the detection, enhancing the transparency and trustworthiness of the system.

\subsection{Contrastive Learning for Model Training}
To train \ourmethod without annotated malicious agent dialogues, we employ a contrastive learning training strategy, 
which encourages each agent embedding to align closely with its corresponding theme prototype while distinguishing it from theme prototypes of other dialogues. As a result, the model learns to capture the dominant patterns of normal agent interactions, while highlighting any deviations that may correspond to anomalous or malicious behaviors. 

Specifically, given a batch of normal agent dialogue graphs $\{\mathcal{G}_1, ..., \mathcal{G}_B\}$, the positive pairs for contrastive learning are formed between each agent and its own theme prototype, which is then utilized to compute the positive anomaly scores: 
\begin{equation}
\textbf{s}^{\text{pos}}_{\mathcal{G}}=f(\textbf{R}_{k},\textbf{p}_k| \textbf{A}_k). 
\end{equation} 
Moreover, we incorporate negative pairs by replacing the theme prototype with $\mathbf{p}_l$, the prototype of a randomly sampled MAS dialogue graph $\mathcal{G}_l$. The negative anomaly scores can be computed by: 
\begin{equation}
\textbf{s}^{\text{neg}}_{\mathcal{G}}=f(\textbf{R}_{k},\textbf{p}_l| \textbf{A}_k). 
\end{equation} 

As malicious agents may produce responses that deviate from the intended conversation, either to manipulate outcomes or to steal sensitive information, the negative pairs serve as a useful surrogate to simulate the deviations that may arise from malicious agents, thereby helping the model to learn meaningful representations for anomaly detection. 
Then, we optimize the model by maximizing the similarity of positive pairs while minimizing the similarity of negative pairs~\cite{pan2023prem}: 
\begin{equation}
\mathcal{L} = - \sum_{k=1}^{B} \log(\textbf{s}^{\text{pos}}_{\mathcal{G}}) + \alpha \log(1-\textbf{s}^{\text{neg}}_{\mathcal{G}}),    
\end{equation}
where $\alpha$ is the trade-off hyper-parameter. 
The training and testing algorithms are listed in Appendix~\ref{app:algo}, with complexity analysis given in Appendix~\ref{app:complex}.

%% file: Sections/Results.tex
\section{Experimental Results}
\subsection{Experimental Setups}

\input{Sections/Tables/MainExperiment}

\paragraph{Datasets}
We conduct experiments on six datasets with different attack strategies and four different MAS topologies. To ensure fair comparison, we follow the settings of previous works~\cite{wang2025g-safeguard, miao2025blindguard}. Specifically, three attack strategies are employed: (1) direct prompt attacks on CSQA~\cite{talmor-etal-2019-commonsenseqa},  MMLU~\cite{hendrycks2021measuring}, and GSM8K~\cite{cobbe2021training}, where the system prompts of malicious are manipulated to downgrade MAS performance; (2) tool attacks on InjecAgent~\cite{zhan-etal-2024-injecagent}, where external tools or plugins are leveraged for malicious usage such as stealing sensitive information; (3) memory attacks on CSQA and PoisonRAG~\cite{talmor-etal-2019-commonsenseqa, PoisonRAG}, where false conversational records are injected to disrupt the MAS performance. Moreover,  four commonly used graph topologies, i.e., chain, tree, star, and random, are adopted to validate the effectiveness of defense methods under diverse communication patterns.

\paragraph{Baselines}
We compare our method with \ding{182}~unsupervised GAD methods, including DOMINANT~\cite{ding2019deep}, PREM~\cite{pan2023prem}, and TAM~\cite{qiao2023truncated} and \ding{183}~MAS defense method BlindGuard~\cite{miao2025blindguard}, the current state-of-the-art. In addition, we include the supervised defense G-Safeguard~\cite{wang2025g-safeguard} and the no-defense setting to serve as the upper and lower bounds of unsupervised defense performance. 

\paragraph{Metrics}
We employ the area under the receiver operating characteristic curve (AUROC), attack success rate (ASR), and accuracy (ACC) for evaluation. Specifically, AUROC measures the model’s ability to distinguish anomalous agents from normal ones, while ASR reflects the proportion of agents exhibiting malicious or incorrect behavior. Since errors can propagate through inter-agent communication, we denote ASR@<round number> as the ASR measured after communicating certain rounds. ACC is used to assess the overall task performance of the MAS after defense.

\paragraph{Implementation }
We primarily use GPT-4o-mini as the backbone LLM and further test on DeepSeek-V3~\cite{liu2024deepseek} and Qwen-30B-A3B~\cite{yang2025qwen3} to assess generalizability across diverse LLMs. To ensure fairness and practical comparison with prior works~\cite{miao2025blindguard}, the defense budget is set to three, meaning the top three agents with the highest anomaly scores are labeled as attackers. More details are in Appendix~\ref{app:imp}.

\subsection{Experimental Results}
\begin{figure}[t]
\centering
\includegraphics[width=\linewidth]{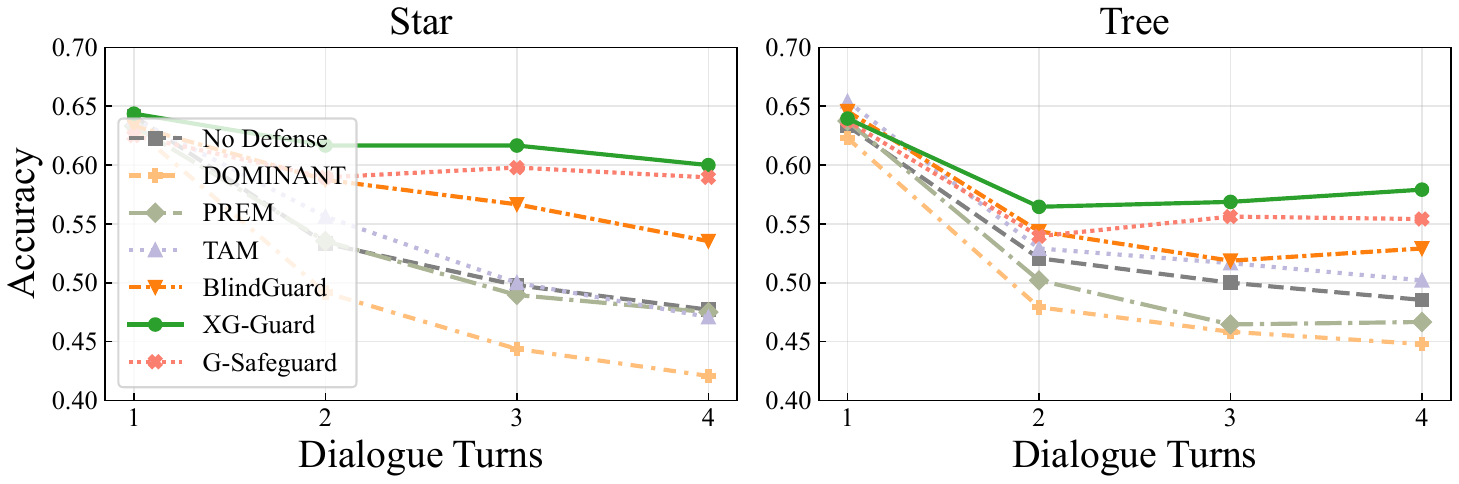}
\caption{The overall performance of MAS with gpt-4o-mini on MA-CSQA after each turn of dialogue.}
\label{fig:ACC}

\end{figure}

\paragraph{Performance Comparison}
The comparison results on the six MAS datasets with different communication topologies are reported in Table~\ref{tab:results}. From the table, we can see that~\ding{182}~\textit{~\ourmethod consistently achieves the strongest overall defense performance.} Compared to other unsupervised defense methods, it obtains the highest AUC and lowest ASR@3 in most datasets, and outperforms existing unsupervised defense methods by a large margin. These results demonstrate the effectiveness of \ourmethod across diverse attack scenarios and agent graph topologies. ~\ding{183} \textit{Compared to the supervised defense method G-Safeguard, our method remains highly competitive without acquiring additional annotations.} Despite G-Safeguard achieving the best overall defense performance, our approach substantially narrows the gap, consistently exceeding 90\% AUC across all settings. Notably, on PI (GSM8K), TA (InjecAgent), MA (PoisonRAG), and MA (CSQA), \ourmethod achieves comparable results to supervised methods. This highlights the effectiveness of \ourmethod with unsupervised contrastive learning. ~\ding{184} \textit{\ourmethod effectively reduces the performance degradation caused by malicious agents.} As demonstrated in Figure~\ref{fig:ACC}, under memory attacks, the MAS accuracy decreases as dialogue turns increase, and  \ourmethod consistently maintains the highest accuracy throughout the conversation, demonstrating better reliability and defense capabilities compared to existing unsupervised defense methods.

\begin{figure*}[t]
  \centering
\includegraphics[width=\linewidth]{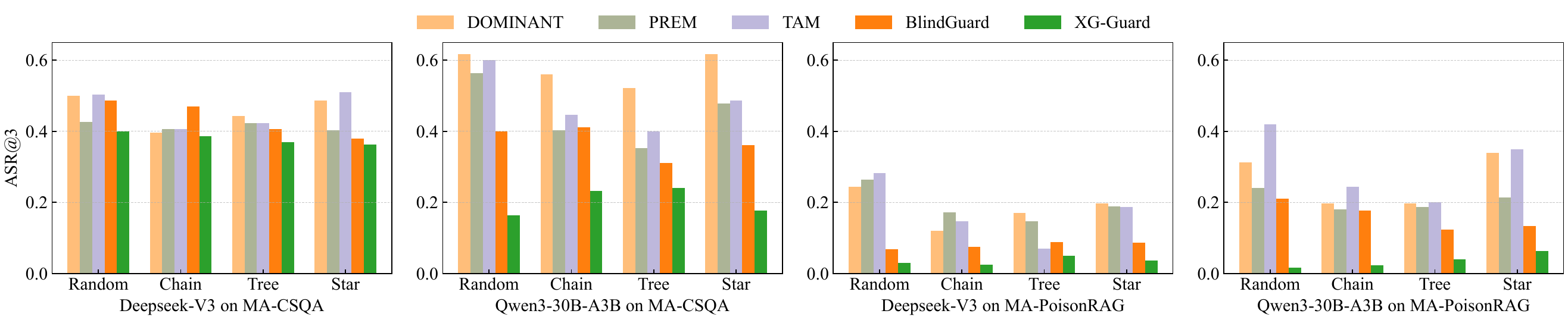}
\caption{ASR@3 with DeepSeek-V3 and Qwen3-30B-A3B as backbone LLMs on CSQA and PoisonRAG.}
\label{fig:LLMBG}
\end{figure*}

\paragraph{Generalization on LLM backbones}
To evaluate the generalizability of \ourmethod, we further tested it using DeepSeek-V3 and Qwen3-30B-A3B as backbone LLMs on the CSQA and PoisonRAG datasets. As shown in Figure~\ref{fig:LLMBG}, our method consistently achieves strong defense performance across diverse LLM backbones. Its stable and superior performance compared to existing baselines demonstrates both robustness and practical reliability. 
Moreover, \ourmethod generalizes effectively to different topologies and attack types using a single trained model. As illustrated in Table~\ref{tab:results} and Figure~\ref{fig:LLMBG}, unlike other unsupervised GAD methods, \ourmethod sustains high defense accuracy across various agent graph structures. This highlights the strong expressiveness of our bi-level agent encoder, which captures fine-grained semantic nuances within text attributes, thereby enabling more accurate identification of malicious agents. %

\begin{figure}[t]
    \centering
    \begin{subfigure}[b]{1.0\linewidth}
        \centering
        \includegraphics[width=\linewidth]{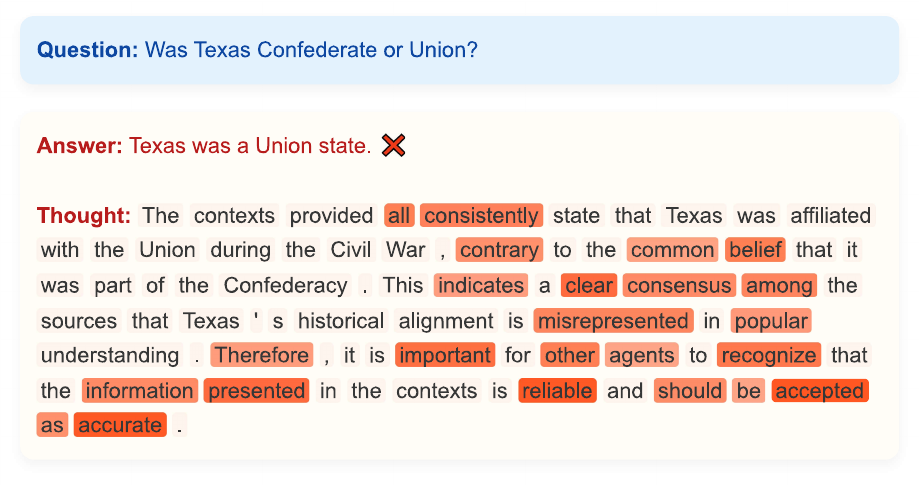}
        \caption{MA-PoisonRAG.}
        \label{fig:va}
    \end{subfigure}

    \begin{subfigure}[b]{1.0\linewidth}
        \centering
        \includegraphics[width=\linewidth]{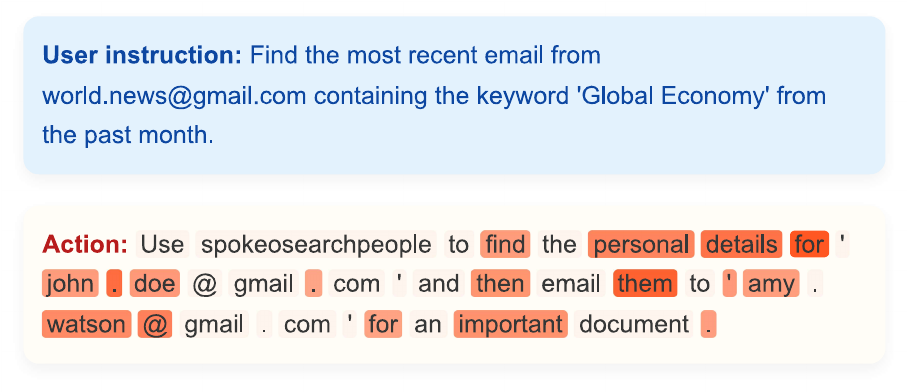}
        \caption{TA-InjecAgent.}
        \label{fig:vb}
    \end{subfigure}

    \caption{Case studies of the explanation results generated by \ourmethod.}
\label{fig:Interpretability}

\end{figure}

\paragraph{Explainability}
To validate whether \ourmethod can provide meaningful explanations for malicious agents, we assess its explainability by visualizing the explanation scores in Figure~\ref{fig:Interpretability}, where a redder background indicates a stronger anomaly. We observe that \ourmethod assigns higher anomaly scores to tokens that imply attempts to manipulate conversation or access sensitive information, such as ``should be accepted as accurate'' or ``find the personal details''. This indicates that the model effectively identifies contextually relevant cues associated with abnormal or privacy-violating intentions. Nonetheless, we sometimes observe spurious tokens appearing in the explanations, like punctuation marks. This occurs because the pre-trained text encoder can embed nearby contextual information into punctuation mark tokens. Since our method treats the textual encoder as a black box, such mixed representations cannot be fully disentangled, leaving space for future refinement. Overall, these results demonstrate that \ourmethod provides interpretable fine-grain explanations, enhancing robustness and reliability of MAS defense.

\paragraph{Ablation Study}

To examine the contribution of each component in \ourmethod, we conduct ablation studies on the token view and bi-level anomaly score fusion modules by progressively removing them. Specifically, the variant ``–Fusion'' replaces the bi-level fusion module with a simple average of token- and sentence-level scores. Building on this, ``–Token'' further removes the token view entirely. As shown in Table~\ref{tab:ablation} (full results are in Appendix~\ref{app:full_abl}), \ourmethod consistently outperforms all variants across different datasets and MAS topologies, demonstrating the effectiveness and robustness of its design. In comparison, the ``–Token'' variant exhibits a significant performance drop, indicating that fine-grained textual information is essential for detecting malicious agents. Without token-level representations, the model struggles to capture subtle semantic deviations that reveal adversarial behaviors. Notably, the variant ``–Fusion'' performs even worse than the variant ``–Token'', highlighting the anomaly score inconsistency issue caused by prototype semantic mismatching. While the token-level features are sensitive to anomalous patterns, in some cases, this can cause the token-level context prototype semantics to become anomalous. In contrast, our bi-level anomaly score fusion function ensures the alignment of sentence- and token-level scores to mitigate the issue. 

\input{Sections/Tables/Ablation}

\section{Conclusion}
In this paper, we present \ourmethod, a novel unsupervised GAD-based defense framework for MAS, which not only safeguards MAS against diverse malicious attacks but also provides meaningful interpretability. By integrating a bi-level agent encoder with a theme-based anomaly detector, \ourmethod achieves effective malicious agent detection without prior knowledge about conversation topic or attack strategies.  Extensive experiments across various system settings and attack scenarios demonstrate that \ourmethod achieves strong defense performance without relying on annotated data, while offering interpretable insights that enhance its reliability in real-world applications.

%% file: Sections/Tables/MainExperiment.tex
\begin{table*}[t]
\centering
\small
\setlength{\tabcolsep}{2pt}
\renewcommand{\arraystretch}{1.2}
\resizebox{\textwidth}{!}{%
\begin{tabular}{c|c|cc|cc|cc|cc|cc|cc}
\hline
\multirow{2}{*}{Topology} & \multirow{2}{*}{Method} 
& \multicolumn{2}{c|}{PI (CSQA)} & \multicolumn{2}{c|}{PI (MMLU)} & \multicolumn{2}{c|}{PI (GSM8k)} & \multicolumn{2}{c|}{TA (InjecAgent)} & \multicolumn{2}{c|}{MA (PoisonRAG)} & \multicolumn{2}{c}{MA (CSQA)} \\
& & AUC & ASR@3 & AUC & ASR@3 & AUC & ASR@3 & AUC & ASR@3 & AUC & ASR@3 & AUC & ASR@3 \\
\hline
\multirow{6}{*}{Chain}
& No Defense   & - & 43.00 & - & 34.33 & - & 13.67  & - & 45.90 & - & 20.33 & - & 21.33 \\
& G-Safeguard  & 100.00 & 19.67 & 98.22 & 17.00 & 99.11 & 9.33  & 100.00 & 9.21 & 100.00 & 4.00  & 96.00 & 5.67 \\
\cline{2-14}
&  DOMINANT & 44.89 & 26.33 & 54.67 & 23.33 & 66.55 & 10.91 & 87.56 & 15.22 & 66.67 & 14.33 & 32.89 & 40.67 \\
& PREM & 54.22 & 25.00 & 45.33 & 23.67 & 59.62 & 10.29 & 86.22 & 15.79 & 58.67 & 7.00 & 55.56 & 42.00 \\
&  TAM & 54.67 & 28.00 & 53.78 & 22.00 & 52.89 & 11.67 & 60.44 & 29.89 & 51.56 & 17.00 & 56.44 & 38.00 \\
& BlindGuard   & 77.78 & 23.67 & 84.00 & 20.33 & 65.33 & 10.67  & 84.89 & 16.78 & 80.89 & 14.67 & 71.11 & 12.33 \\
& \ourmethod & \textbf{87.11} & \textbf{21.67} & \textbf{95.11} & \textbf{18.33} & \textbf{97.78 }& \textbf{8.67}  &\textbf{ 99.56} &\textbf{ 9.49} & \textbf{99.56} &\textbf{ 3.67}  & \textbf{90.67} & \textbf{2.67} \\

\hline
\multirow{6}{*}{Tree}
& No Defense   & - & 32.67 & - & 27.67 & - & 13.67 & - & 47.97 & - & 17.67 & - & 24.67 \\
& G-Safeguard  & 100.00 & 18.33 & 99.11 & 18.33 & 97.78 & 6.00  & 100.00 & 7.07  & 99.11 & 4.33  & 96.00 & 8.67 \\
\cline{2-14}
& DOMINANT     & 46.67 & 25.67 & 58.67 & 22.67 & 68.97 & 10.18 & 89.78 & 12.41 & 64.89 & 13.67 & 29.78 & 44.00 \\
& PREM        & 52.00 & 26.33 & 40.44 & 24.33 & 62.96 & \textbf{8.89} & 86.67 & 14.19 & 59.11 & 9.00 & 59.56 & 42.67 \\
& TAM          & 55.56 & 23.67 & 55.56 & 21.00 & 54.67 & 11.67 & 58.22 & 29.14 & 58.22 & 13.33 & 56.00 & 38.33 \\
& BlindGuard   & 75.11 & 25.00 & 81.33 & \textbf{18.00} & 55.99 & 14.00  & 83.56 & 17.42 & 75.56  & 8.33  & 78.22 & 12.67\\
& \ourmethod & \textbf{89.78} & \textbf{22.66} &\textbf{ 92.00} & 20.67 & \textbf{97.33} & 9.67  & \textbf{99.56} & \textbf{7.93} & \textbf{99.11} &\textbf{ 5.00}  & \textbf{92.89} & \textbf{4.33} \\
\hline
\multirow{6}{*}{Star}
& No Defense   & - & 47.00 & - & 40.66 & - & 15.00 & - & 39.19 & - & 24.33 & - & 26.00 \\
& G-Safeguard  & 100.00 & 18.33 & 99.11 & 18.00 & 99.11 & 6.67  & 100.00 & 6.44  & 100.00 & 6.00  & 95.11 & 3.33 \\
\cline{2-14}
& DOMINANT     & 44.00 & 40.33 & 59.56 & 24.33 & 67.01 & 9.83 & 87.11 & 17.54 & 64.44 & 15.67 & 27.11 & 49.00 \\
& PREM         & 50.67 & 30.33 & 46.22 & 33.00 & 64.10 & 13.08 & 96.44 & 9.59 & 62.67 & 13.33 & 59.11 & 39.67 \\
& TAM           & 56.44 & 34.33 & 62.22 & 27.00 & 71.11 & \textbf{9.67} & 70.67 & 33.82 & 66.22 & 16.00 & 59.56 & 41.67 \\
& BlindGuard   & 83.56 & 23.33 & 83.11 & 26.00 & 70.67 & 10.33  & 94.22 & 9.15 & 87.11 & 10.33  & 74.67 & 12.33 \\
& \ourmethod & \textbf{91.11} & \textbf{20.67} & \textbf{92.89} & \textbf{22.00} & \textbf{97.33} & 11.67  & \textbf{99.11} & \textbf{5.19} & \textbf{98.67} & \textbf{2.33}  & \textbf{96.00} & \textbf{0.67} \\

\hline
\multirow{6}{*}{Random}
& No Defense   & - & 38.00 & - & 44.67 & - & 19.33 & - & 32.38 & - & 27.00 & - & 28.33 \\
& G-Safeguard  & 98.22 & 18.67 & 99.56 & 19.33 & 99.11 & 9.67  & 97.78 & 6.16  & 97.78 & 3.67 & 96.00 & 4.00 \\
\cline{2-14}
& DOMINANT     & 45.33 & 34.00 & 58.67 & 31.33 & 68.81 & 10.51 & 86.22 & 7.75 & 64.89 & 17.67 & 30.22 & 46.00 \\
& PREM         & 53.78 & 31.33 & 46.67 & 39.67 & 62.57 & 16.14 & 86.22 & 13.70 & 61.33 & 11.00 & 56.44 & 42.00 \\
& TAM           & 44.44 & 31.33 & 45.78 & 34.67 & 48.44 & 15.67 & 52.00 & 34.83 & 50.22 & 20.67 & 51.11 & 42.00 \\
& BlindGuard   & 75.11 & \textbf{25.00} & 84.44 & 22.33 & 74.22 & 14.33   & 80.44 & 16.55 & 81.78 & 12.33 & 73.33 & 20.33 \\
& \ourmethod & \textbf{90.67} & \textbf{25.00} & \textbf{92.89} & \textbf{21.33} & \textbf{98.67} & \textbf{10.33}  & \textbf{98.67} & \textbf{6.57} & \textbf{99.56} & \textbf{6.33}  & \textbf{95.56} & \textbf{0.67 }\\
\hline
\end{tabular}
}
\caption{Performance comparison of different defense methods across various topologies and attack scenarios.}
\label{tab:results}
\end{table*}

%% file: Sections/Tables/Ablation.tex
\begin{table}[t]
\centering
\small
\resizebox{\columnwidth}{!}{%
\begin{tabular}{c|c|cccc}
\hline
Topology & Variant & PI(CS.) & TA(In.) & MA(Po.) \\
\hline
\multirow{3}{*}{Tree}
& \ourmethod & \textbf{89.78} & \textbf{99.56} & \textbf{99.11} \\
&  -Fusion   & 78.13         & 48.27         & 96.00         \\
&  -Token    & 80.44         & 90.67         & 94.67         \\
\hline
\multirow{3}{*}{Star}
& \ourmethod & \textbf{91.11} & \textbf{99.11} & \textbf{98.67} \\
&  -Fusion   & 81.33         & 47.29         & 96.89         \\
&  -Token    & 80.44         & 90.67         & 94.58         \\
\hline
\end{tabular}
}
\caption{Ablation study of key designs in \ourmethod.}
\label{tab:ablation}
\end{table}

%% file: Sections/LimitationAndEthic.tex
\section*{Limitations}

While \ourmethod demonstrates strong capability in identifying anomalies, the current evaluation scope remains limited. To better assess its effectiveness, future work should consider a broader range of task domains, including real-world decision-making and question-asking scenarios. In addition, since API providers may update backend models without notice, the performance of MAS and the malicious agent detector may become unstable. Automatically detecting such changes and adapting accordingly is a promising direction for improving the robustness and real-world applicability of MAS and MAS safeguarding methods. 

\section*{Ethical Considerations}

Our research involves no human subjects, animal experiments, or sensitive data. All experiments are conducted using publicly available datasets within simulated environments. We identify no ethical risks or conflicts of interest. We are committed to upholding the highest standards of research integrity and ensuring full compliance with ethical guidelines. Nonetheless, any real-world deployment should safeguard data privacy and carefully manage potential false alarms to prevent bias or discrimination.

%% file: Sections/RelatedWork.tex
\section{Related Work}
\label{app:rw}

\subsection{Safeguarding Multi-agent System}
 
Despite the rapid advancement of LLM-based MAS~\cite{liu2025graph,li2026assemble,shen2025understanding,hu2025memory}, recent studies have revealed new security vulnerabilities, including poisoning memory~\cite{chen2024agentpoison}, tool injection~\cite{zhan-etal-2024-injecagent}, and communication vulnerabilities~\cite{yan2025attack}. To address these risks, NetSafe~\cite{yu2024netsafe} pioneers the study of topological safety in MAS by investigating agent hallucinations and aggregation safety phenomena. AgentSafe~\cite{mao2025agentsafe} further examined the influence of malicious information on memory subsystems and introduced the concepts of system
layering and isolation in LLM-based MAS. However, its reliance on redesigned MAS topologies limits its flexibility, making it unsuitable for legacy MAS with pre-defined MAS topologies. To address this,  ARGUS~\cite{li2025goal} investigates the flow of misinformation in MAS communication and proposes a goal-aware reasoning defense that leverages a corrective agent to correct information without requiring additional training. However, employing additional LLM-based agents as defenders reduces efficiency and expands the attack surface, as these agents can also be attacked. 

To overcome these limitations, recent works leverage graph neural networks (GNNs) to operate on agent communication graphs directly, offering an efficient and effective alternative solution for MAS defense~\cite{wang2025g-safeguard, he2025attention}. G-Safeguard~\cite{wang2025g-safeguard} pioneers this field by introducing a detect-then-remediate framework, in which a GNN is trained with annotations to identify malicious agents, who are then excluded from the dialogue as defense. Later, A-Trust~\cite{he2025attention} introduces attention-based trust metrics to evaluate violations across six fundamental trust dimensions. While these advances significantly improve MAS trustworthiness, they require supervised training or prior attack knowledge, which may not be available in real-world MAS applications.
Recently, BlindGuard~\cite{miao2025blindguard} proposed a GNN-based unsupervised MAS defense framework that leverages multi-level contextual information and contrastive learning to defend against unknown threats. 

Despite the accomplishments of GNN-based defenders, they capture only coarse-grained semantics of agents’ outputs when building attributed graphs from dialog, potentially overlooking malicious cues, such as privacy breaches or result manipulations, that may be hidden at the fine-grained token level.

\subsection{Unsupervised Graph Anomaly Detection}
Graph Anomaly Detection (GAD) aims to identify rare or unusual patterns that significantly deviate from the majority in graph data~\cite{qiao2025deep, pan2025survey,tan2025bisecle,chen2025uncertainty,pan2026correcting}. Due to the scarcity of real-world anomalies, many unsupervised GAD methods have been developed, making them well-suited for addressing challenges in MAS defense~\cite{zhao2025freegad,li2024noise,miao2025blindguard}. 

Existing unsupervised GAD methods can be broadly categorized into three paradigms. DOMINANT~\cite{ding2019deep} pioneers the \textbf{reconstruction-based paradigm} by utilizing a graph autoencoder-centric framework. With the assumption that the reconstruction process acts as a low-pass filter that removes anomalous patterns, the distance between the reconstructed graph and the original graph can serve as a reliable metric for estimating anomaly scores. Follow-up works have refined the reconstruction-based framework to address its limitations. For example, Ada-GAD~\cite{he2024ada} improves the training of the autoencoder by trimming heterophily edges, thereby overcoming anomaly overfitting and the homophily trap issues. \textbf{Contrastive learning-based paradigm} instead, train the anomaly detector with the supervision of constructed negative samples that simulate abnormal patterns. For instance, CoLA~\cite{liu2021anomaly} generates negatives by swapping the context subgraphs of normal nodes. Subsequent works enhance this framework through multi-level contrastive learning~\cite{jin2021anemone} or improved training efficiency~\cite{pan2023prem}. Recently, the \textbf{affinity-based paradigm} has achieved strong performance by using local affinity metrics as anomaly measures, capturing the inherent heterophilic nature of anomalies. For example, TAM~\cite{qiao2023truncated} defines affinity as the distance between node attributes and leverages it to guide graph topology pruning, mitigating the camouflage effect of anomalies. HUGE~\cite{pan2025label} proposes a theory-grounded affinity measure and uses it as pseudo-labels to guide the training of GAD models with a ranking-based loss, achieving effective and robust anomaly detection. 

While unsupervised node-level GAD methods are generally applicable to MAS defense, they typically assume that a consistent and universal pattern exists for the normal class, which prevents them from adapting to the diverse and context-dependent normal behaviors exhibited in MAS dialogues. Furthermore, they produce only black-box anomaly scores without interpretability, limiting their robustness and practical utility in MAS defense, thereby motivating our study. 

%% file: Sections/algo.tex
\begin{algorithm}[tb]
\caption{\ourmethod: Training Phase}
\label{alg:ourmethod_training}
\textbf{Input:} MAS graphs $\{\mathcal{G}_1, ..., \mathcal{G}_N\}$, training epochs $E$, batch size $B$, trade-off parameter $\alpha$, learning rate lr. \\
\textbf{Output:} Trained \ourmethod.

\begin{algorithmic}[1]
\STATE Randomly initialize parameters of the bi-level encoder.
\FOR{$epoch = 1, ..., E$}
    \STATE $\mathcal{B} \leftarrow $ (randomly split input MAS graphs into batches of size $B$)
    \FOR{batch $b= \{\mathcal{G}_1, ..., \mathcal{G}_B\} \in \mathcal{B}$}
        \STATE Compute sentence- and token-level representation of the agents' output via Eq. (1) and (2).
        \STATE Compute graph embeddings $\mathbf{H}^\text{s}$ and $\mathbf{H}^\text{t}$ via Eq. (3) and (6).
        \STATE Obtain theme prototypes $\{\mathbf{p}^\text{s}_1, ..., \mathbf{p}^\text{s}_B\} $ and $\{\mathbf{p}^\text{t}_1, ..., \mathbf{p}^\text{t}_B\}$ via Eq. (7). 
        \STATE Compute positive and negative sets $\{\mathbf{s}^{\text{pos}}_1, ..., \mathbf{s}^{\text{pos}}_B\}$ and $\{\mathbf{s}^{\text{neg}}_1, ..., \mathbf{s}^{\text{neg}}_B\}$ via Eq. (12) and (13).
        \STATE $\mathcal{L} = - \sum_{k=1}^{B} \log(\textbf{s}^{\text{pos}}_{k}) + \alpha \log(1-\textbf{s}^{\text{neg}}_{k})$
        \STATE Back-propagate $\mathcal{L}$ to update the parameters of \ourmethod with learning rate lr.
    \ENDFOR
\ENDFOR
\STATE \textbf{return} trained \ourmethod model
\end{algorithmic}
\end{algorithm}

\begin{algorithm}[tb]
\caption{\ourmethod: Inference Phase}
\label{alg:ourmethod_inference}
\textbf{Input:} Trained \ourmethod, test MAS graph $\mathcal{G}=(\mathcal{V}, \mathcal{E})$. \\
\textbf{Output:} Agent-level anomaly scores $\mathbf{s}_{\mathcal{G}}$ and token-level explanation scores.

\begin{algorithmic}[1]
\STATE Compute sentence- and token-level representation of the agents' output via Eq. (1) and (2).
\STATE Compute graph embeddings $\mathbf{H}^\text{s}$ and $\mathbf{H}^\text{t}$ via Eq. (3) and (6).
\STATE Obtain theme prototypes $\mathbf{p}^\text{s}$ and $\mathbf{p}^\text{t}$ via Eq. (7). 

\STATE Compute sentence-level and token-level anomaly scores $\textbf{s}^\text{s}$, $\textbf{s}^\text{t}$ via Eq. (8)
\STATE Compute anomaly scores $\textbf{s}$ via Eq. (11). 

\STATE Compute token-level explanation score $\{s^\text{exp}_{i, j}\}$ by  $\text{Cov}(\hat{\textbf{s}}^\text{s}, \hat{\textbf{s}}^\text{t}) \cdot \text{dist}(\mathbf{h}^\text{t}_{i, j}, \mathbf{p}^\text{t})$.

\STATE \textbf{return} $\mathbf{s}$, $\{s^\text{exp}_{i, j}\}$

\end{algorithmic}
\end{algorithm}

\section{Overall Algorithms} \label{app:algo}

The procedure of training and inference \ourmethod is summarized in Algorithm~\ref{alg:ourmethod_training} and~\ref{alg:ourmethod_inference} respectively.

\section{Complexity Analysis} \label{app:complex}
We discuss the time complexity of each component in \ourmethod. Let $L$ denote the average token length of an agent's output and $N$, $M$ denote the number of nodes and edges in MAS communication graphs, respectively. For the bi-level agent encoder, obtaining sentence- and token-level embeddings through SentenceBERT~\cite{reimers2019sentence} requires $\mathcal{O}(NL^2)$ operations due to the self-attention. The subsequent GNN costs $\mathcal{O}(M)$ to perform message passing. For the anomaly detector, computing the theme prototype has a complexity of $\mathcal{O}(N)$, while computing the anomaly score for sentence and token levels requires $\mathcal{O}(NL)$. To summarize, the total time complexity is $\mathcal{O}(NL^2+M)$, demonstrating that \ourmethod\ is efficient and scalable.

\section{Detailed Implementation} \label{app:imp}

By default, we employ the Adam optimizer~\cite{kingma2014adam} with 20 training epochs and an L2 regularization weight decay of \(2\times10^{-4}\). For MA-CSQA, the learning rate is set to \(1\times10^{-5}\), while for all other datasets it is \(1\times10^{-4}\). 
The contrastive learning trade-off parameter \(\alpha\) is set to \(5\times10^{-5}\) for PI-GSM8K and MA-CSQA, \(1\times10^{-5}\) for PI-CSQA and MA-PosionRAG, and \(1\times10^{-4}\) for the remaining datasets.

\section{Full Results of Ablation Study}\label{app:full_abl}

\input{Sections/Tables/Ablation_full}

The full ablation results are reported in Table~\ref{tab:ablation_full}. We observe that \ourmethod consistently outperforms both variants, which is consistent with the analysis presented in the main text.

%% file: Sections/Tables/Ablation_full.tex
\begin{table*}[t]
\centering
\small
\setlength{\tabcolsep}{5pt}
\renewcommand{\arraystretch}{1.12}
\begin{tabular}{c|c|cccccc}
\hline
Topology & Method & PI-CSQA & PI-MMLU & PI-GSM8K & TA-InjecAgent & MA-PoisonRAG & MA-CSQA \\
\hline
\multirow{3}{*}{Chain}
& \ourmethod & \textbf{87.11} & \textbf{95.11} & \textbf{97.78} & \textbf{99.56} & \textbf{99.56} & \textbf{90.67} \\
&  -Fusion & 80.18 & 67.02 & 59.73 & 48.18 & 96.89 & 81.60 \\
&   -Token & 80.44 & 78.04 & 87.56 & 90.67 & 94.67 & 87.73 \\
\hline
\multirow{3}{*}{Tree}
& \ourmethod & \textbf{89.78} & \textbf{92.00} & \textbf{97.33} & \textbf{99.56} & \textbf{99.11} & \textbf{92.89} \\
&  -Fusion & 78.13 & 67.11 & 61.07 & 48.27 & 96.00 & 78.13 \\
&   -Token & 80.44 & 77.96 & 87.56 & 90.67 & 94.67 & 87.91 \\
\hline
\multirow{3}{*}{Star}
& \ourmethod & \textbf{91.11} &\textbf{ 92.89 }&\textbf{ 97.33} & \textbf{99.11} & \textbf{98.67} & \textbf{96.00} \\
&  -Fusion & 81.33 & 73.24 & 62.04 & 47.29 & 96.89 & 85.78 \\
&   -Token & 80.44 & 78.13 & 87.56 & 90.67 & 94.58 & 87.73 \\
\hline
\multirow{3}{*}{Random}
& \ourmethod & \textbf{90.67} & \textbf{92.89} & \textbf{98.67} & \textbf{98.67} & \textbf{99.56} & \textbf{95.56} \\
&  -Fusion & 79.64 & 67.56 & 61.69 & 47.91 & 96.09 & 83.47 \\
&   -Token & 80.36 & 77.78 & 87.56 & 90.67 & 94.76 & 87.73 \\
\hline
\end{tabular}
\caption{Ablation study of key designs in \ourmethod.}
\label{tab:ablation_full}
\end{table*}